\documentclass[twocolumn,aps,superscriptaddress,nofootinbib,floatfix,linenumbers]{revtex4}
\usepackage{epsfig,bm,feynmf}
\usepackage{graphics}
\usepackage{amsmath}
\usepackage{mathrsfs}
\usepackage{appendix}
\usepackage{bm}
\usepackage[normalem]{ulem}  
\usepackage[dvips]{color} 

\renewcommand{\sout}{\bgroup \color{red} \ULdepth=-.5ex \ULset}

\begin{document}
\title{Study of chiral vortical and magnetic effects in the anomalous transport model}


\author{Yifeng Sun}
\email{sunyfphy@physics.tamu.edu}
\affiliation{Cyclotron Institute and Department of Physics and Astronomy, Texas A$\&$M University, College Station, Texas 77843, USA}%

\author{Che Ming Ko}
\email{ko@comp.tamu.edu}
\affiliation{Cyclotron Institute and Department of Physics and Astronomy, Texas A$\&$M University, College Station, Texas 77843, USA}%

\date{\today}

\begin{abstract}
We extend our recent study of chiral magnetic effect in relativistic heavy ion collisions based on an anomalous transport model by including also the chiral vortical effect. We find that although vorticities in the chirally restored quark matter, which result from the large angular momentum in non-central collisions, can generate an axial charge dipole moment
in the transverse plane of a heavy ion collision, it does not produce a difference in the eccentricities of negatively and positively charged particles.  As a result, including the chiral vortical effect alone cannot lead to a splitting between the elliptic flows of negatively and positively charged particles.  On the other hand, negatively and positively charged particles do develop a splitting in their elliptic flows if the effect due to a strong and long-lived magnetic field is also included. However, to have a positive slope in the dependence of the elliptic flow splitting on the charge asymmetry of the quark matter, as seen in experiments, requires the neglect of the effect of the Lorentz force. In this case, an elliptic flow splitting appears even at vanishing charge asymmetry.
\end{abstract}

\maketitle

\section{introduction}

Novel transport phenomena induced by the chiral anomaly have recently generated a lot of interests in various areas of physics~\cite{Charbonneau:2009ax,PhysRevB.86.115133,PhysRevB.89.035142}. One such phenomenon is the chiral magnetic (separation) effect (CME/CSE) in which vector (axial) charges of massless fermions become spatially separated in the presence of a magnetic field if the axial (vector) charge chemical potential is nonzero~\cite{PhysRevD.70.074018,PhysRevD.72.045011,Kharzeev2008227,PhysRevD.78.074033,Kharzeev2010205}. Another is the chiral vortical effect (CVE) in which a vector current appears in a rotating system when both the vector and axial chemical potentials are nonzero~\cite{PhysRevLett.103.191601}. Moreover, through the interplay between the vector and axial charge fluctuations, two gapless collective excitations called the chiral magnetic wave (CMW) and the chiral vortical wave (CVM) can be induced~\cite{PhysRevLett.107.052303,PhysRevD.92.071501}.

In relativistic heavy ion collisions, a quark-gluon plasma (QGP) consisting of de-confined quarks and gluons can be produced. Because of the restoration of chiral symmetry in the QGP, quarks and antiquarks become essentially massless. Non-central heavy ion collisions, in which both a strong magnetic field and a large vorticity are produced in the QGP as a result of the large initial orbital angular momentum~\cite{PhysRevC.83.054911,PhysRevC.85.044907}, thus provide the opportunity to study above mentioned phenomena.  By solving the wave equation for the charge densities of particles of right- and left-chiralities in a schematic model~\cite{PhysRevLett.107.052303,PhysRevD.92.071501}, it has been found that an electric or baryon charge quadrupole moment is produced, which can then lead to a splitting between the elliptic flows of positively and negatively charged particles as observed in experiments.  More realistic studies of CMW and CVW in heavy ion collisions have been carried out using the anomalous hydrodynamics, which extends the normal hydrodynamics by including also the axial charge current in QGP~\cite{PhysRevLett.103.191601}.  Also, chiral kinetic equations, which include the anomalous transport of massless fermions and the non-equilibrium effect, have been developed~\cite{PhysRevLett.109.162001,Son:2012wh,Son:2012zy,PhysRevLett.109.232301,PhysRevLett.110.262301,Manuel:2014dza}. However, different methods have been proposed in including the chiral vortical effect in the chiral kinetic approach. For example, in Refs.~\cite{PhysRevLett.109.232301,PhysRevLett.110.262301}, the vorticity effect is included in the equations of motion for massless fermions, while in Refs.~\cite{PhysRevLett.113.182302,PhysRevLett.115.021601}, the chiral vortical effect is taken into account via the jump current resulting from the side jump that is introduced in the scattering of massless fermions to conserve the angular momentum.

Using the anomalous transport model for massless quarks and antiquarks, we have recently studied the effect of a magnetic field on the elliptic flows of quarks and antiquarks in relativistic heavy ion collisions~\cite{PhysRevC.94.045204} . With initial conditions from a blast wave model and assuming that the strong magnetic field produced in non-central heavy ion collisions can last for a sufficiently long time, we have obtained an appreciable electric quadrupole moment in the transverse plane of a heavy ion collision. The electric quadrupole moment subsequently leads to a splitting between the elliptic flows of quarks and antiquarks during the expansion of the system. The slope in the charge asymmetry dependence of the elliptic flow difference between negatively and positively charged particles is found to be positive as expected from the chiral magnetic wave in the QGP and observed in experiments at the BNL Relativistic Heavy Ion Collider (RHIC). This result is, however, obtained only if we neglect the Lorentz force acting on the charged particles and assume that the quark-antiquark scattering is dominated by the chirality-changing channel. In the present study, we extend our previous study by including also the vorticities in the partonic matter and investigate how our previous results, particularly the splitting between the elliptic flows of negatively and positively charged particles, are affected.

This paper is organized as follows. In the next section, we describe the chiral equations of motion that include both effects of the magnetic field and the finite vorticity in the QGP. We then give in Sec. III the initial conditions for the partonic matter produced in non-central heavy ion collisions at the top energy of RHIC.  In Sec. IV, we present and discuss the results. Finally, a summary is given in Sec. V.

\section{The chiral equations of motion}

For the chiral kinetic equation that includes both the magnetic and vorticity fields, we follow that obtained in Ref.~\cite{PhysRevLett.110.262301} using the covariant Wigner function approach for massless spin-1/2 fermions in four dimensions.  After integrating over the energy in the Lorentz covariant chiral kinetic equation, a chiral kinetic equation in three dimensions is obtained. The resulting chiral equations of motion for massless quarks and antiquarks are given by
\begin{eqnarray}
&&\sqrt{G}\dot{\mathbf{r}}=\hat{\mathbf{p}}+Qh(\hat{\mathbf{p}}\cdot\mathbf{b})\mathbf{B}+h\frac{\boldsymbol{\omega}}{p},
\label{CKM}
\\&&\sqrt{G}\dot{\mathbf{p}}=Q\hat{\mathbf{p}}\times\mathbf{B},
\label{LF}
\\&&\sqrt{G}=1+Qh\mathbf{b}\cdot\mathbf{B}+4hp(\mathbf{b}\cdot\boldsymbol{\omega}).
\label{phase}
\end{eqnarray}
In the above, $\mathbf{b}=\frac{\mathbf{p}}{2p^3}$ is the Berry curvature due to a vector potential in the momentum space, $h$ is the helicity of the particle, and $\boldsymbol{\omega}=\frac{1} {2}\boldsymbol{ \nabla}\times\mathbf{u}$ is the vorticity with ${\bf u}$ being the velocity field.

As shown in our previous study of CMW based on the anomalous transport model~\cite{PhysRevC.94.045204}, the chirality-changing scattering (CCS) between quark and antiquark is essential for generating the elliptic flow difference between negatively and positively charged particles. We thus also allow in the present study massless quarks and antiquarks of same chiralities to undergo the CCS scattering.  We neglect, however, the effect from the change of phase space, i.e., Eq.(\ref{phase}), due to the magnetic field and/or the vorticity field as in Ref.~\cite{PhysRevC.94.045204}.

\section{Initial conditions of non-central heavy ion collisions}

For the initial conditions of a heavy ion collision, we use same ones in our previous study~\cite{PhysRevC.94.045204}, i.e., a fireball with geometry corresponding to collisions of Au+Au at $\sqrt{s_{NN}}=200$ GeV at centralities of 30-40$\%$.  Specifically, the number density distribution is taken to have the Woods-Saxon form
\begin{eqnarray}
\rho(x,y)=\frac{\rho_0}{1+e^{\frac{\sqrt{x^2+y^2/c^2}-R}{a}}},
\end{eqnarray}
where $\rho_0=13$ fm$^{-3}$ is the central density of quarks and antiquarks, $c=1.5$ describes the spatial anisotropy of produced partonic matter in the transverse plane of non-central heavy ion collisions, $R=3.5$ fm is the radius, and  $a=0.5$ fm is the surface thickness.  The transverse momentum distributions of quarks and antiquarks are given by the Boltzmann distributions of temperature $T$ and charge chemical potential $\mu$ with $\mu/T$ being uniform in space as in
our previous study.  We determine the temperature from the density of quarks and antiquarks by assuming they are noninteracting and the charge chemical potential via $\mu/T=A_\pm$, where $A_\pm=\frac{N_{+}-N_{-}}{N_{+}+N_{-}}$ is the charge asymmetry of the partonic matter with $N_+$ and $N_-$ being the numbers of positively and negatively charged quarks, respectively.

For both the magnetic field and the vorticity field produced in these collisions, we take their directions along the $y$-axis perpendicular to the reaction plane, and assume that their strengths are uniform in space~\cite{PhysRevLett.109.202303} and have the following time dependence:
\begin{eqnarray}
eB&=&\frac{eB_0}{1+(t/\tau_B)^2}\\
\omega&=&\frac{\omega_0}{1+(t/\tau_{\omega})^2}.
\end{eqnarray}
We choose $eB_0=7m_{\pi}^2$ and $\tau_B=6$ fm/$c$ for the magnetic field as in Ref.~\cite{PhysRevC.94.045204}, and $\omega_0=0.1$ fm$^{-1}$ and $\tau_{\omega}=2.7$ fm/$c$ for the vorticity field. The latter is similar to those of Ref.~\cite{PhysRevC.94.044910} based on a multipase tranport (AMPT) model~\cite{Lin:2004en} for Au+Au collision at 200 GeV and centrality 30-40$\%$, except that we take the vorticity field in heavy ion collisions as a spatially uniform external field.

In modeling the time evolution of the partonic matter, massless quarks and antiquarks not only follow the chiral equation of motion [Eq. (\ref{CKM})] and subject to the Lorentz force [Eq. (\ref{LF})] but also undergo scattering, particularly the chirality-changing scattering between quark and antiquark of same chirality. As in our previous study, we take these scattering cross sections to have the temperature dependence $\sigma=\sigma_0(T_0/T)^3$~\cite{Ghosh:2015mda} in order to model the specific viscosity $\eta/s$ of the QGP, where $\eta$ and $s$ are the viscosity and entropy, and its temperature dependence obtained from the lattice QCD and experiments~\cite{Heinz:2013th}.  However, when the temperature of the partonic matter drops below the phase transition temperature $T_C=150$ MeV, we adopt the duality ansatz as in our previous study~\cite{PhysRevC.94.045204} by relabeling each parton as a pion of same sign in charge and evolve the resulting pionic matter using the normal equations of motion and the empirical pion-pion scattering cross sections~\cite{Li:1995pra} until the kinetic freeze out temperature $T_f=120$ MeV.

\section{Results}

In the present section, we study the time evolution of the partonic matter by following the trajectories of massless quarks and antiquarks according to the chiral equations of equations [Eqs. (\ref{CKM}) and (\ref{LF})]. We consider the three cases of including only the vorticity field (CVE), both magnetic and vorticity fields without the Lorentz force (CVE+CME), and both  magnetic field and vorticity fields with the Lorentz force (CVE+CME+LF).  In all cases, the chirality-changing scattering (CCS) between quark and antiquark of same chirality is included, although it has no effect in the case of CVE only.

\subsection{Pion elliptic flow}

\begin{figure}[h]
\centering
\includegraphics[width=0.5\textwidth]{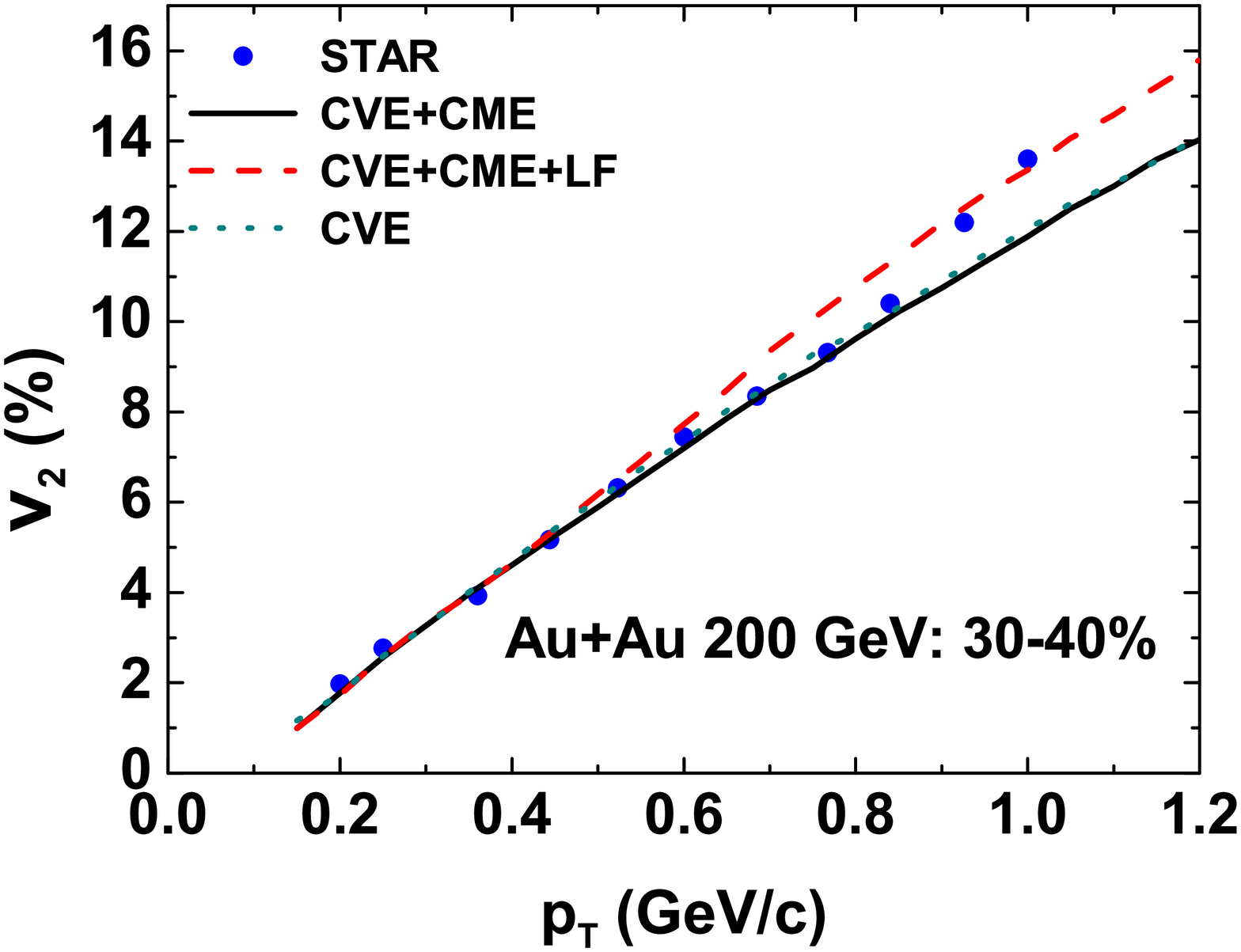}
\caption{(Color online) Elliptic flow of kinetically freeze-out pions  in midrapidity ($|y|\le1$) as a function of transverse momentum for the three cases of including both chiral vortical and magnetic effects and the effect due to the Lorentz force (CVE+CME+LF), without the Lorentz force (CVE+CME), and with only the vortical effect (CVE).  Experimental data (solid circles) are from Ref.~\cite{PhysRevC.72.014904}.}
\label{pion}
\end{figure}

We first show in Fig. \ref{pion} the elliptic flow of kinetically freeze-out pions in midrapidity ($|y|\le1$) as a function of transverse momentum. The dashed line is obtained with the coefficient $\sigma_0=13.6$ mb in the parton scattering cross section for the case of including the chiral effects due to both the vorticity and magnetic fields as well as the Lorentz force (CVE+CME+LF).  The result obtained without the Lorentz force (CVE+CME) and using $\sigma_0=15.5$ mb is shown by the solid line, while the result including only the chiral vortical effect (CVE) and using $\sigma_0=13.3$ mb is shown by the dotted line. The transverse momentum dependence of the pion elliptic flow is seen to agree with the experimental data from the STAR Collaboration~\cite{PhysRevC.72.014904} for all three cases. The integrated $v_2$ of pions with transverse momenta in the range of $0.15\le p_T\le0.5$ GeV/$c$ is 0.036 in all three cases, which also agrees with the experimental value.

\subsection{Effect of the vorticity field}

In Ref.~\cite{PhysRevD.92.071501}, it is argued that the chiral vortical wave (CVW) generated in a rotating fluid system can lead to a splitting between the elliptic flows of baryons and anti-baryons if the baryon chemical potential of the system is non-zero. Similarly, for non-zero charge chemical potential, we expect the CVW to lead to a splitting between the elliptic flows of negatively and positively charged particles. In this section, we use the anomalous transport model to study the effect of the vorticity field on the elliptic flow splitting of $\pi^+$ and $\pi^-$ for different values of charge chemical potential by taking the magnetic field to be zero, i.e.,  $B_0=0$.

\begin{figure}[h]
\centering
\includegraphics[width=0.5\textwidth]{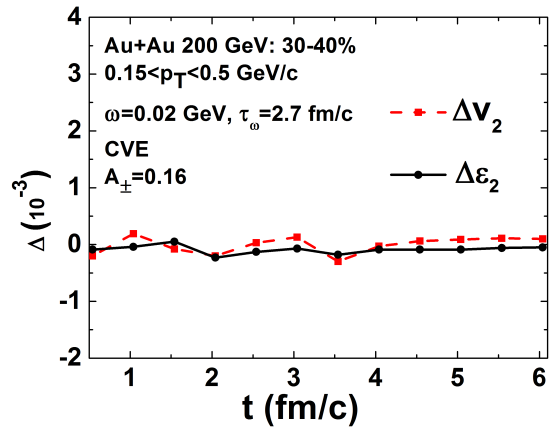}
\caption{(Color online) Eccentricity and elliptic flow difference between negatively and positively charged particles as functions of time for total charge asymmetry of quark matter $A_{\pm}=0.16$.}
\label{cvw}
\end{figure}

Figure \ref{cvw} shows the time evolution of the differences $\Delta \epsilon_2=\epsilon_{2-}-\epsilon_{2+}$ between the eccentricities [$\epsilon_2=\langle (x^2-y^2)/(x^2+y^2)\rangle$] and $\Delta v_2=v_{2-}-v_{2+}$ between the elliptic flows [$v_2=\langle (p_x^2-p_y^2)/(p_x^2+p_y^2)\rangle$] of negatively and positively charged particles for the total charge asymmetry $A_{\pm}=0.16$ of these particles. The transverse momenta are again in the range of $0.15\le p_T\le 0.5$ GeV/$c$. These differences are consistent with zero, indicating that the CVW alone does not lead to any eccentricity and elliptic flow differences between charged particles.

The above result can be understood as follows.  According to the chiral equation of motion [Eqs. (\ref{CKM})] for the case without magnetic field, the vorticity field modifies the velocities of particles of opposite helicities by the same amount but with opposite sign.  As a result, nonzero positive and negative axial charge chemical potentials $\mu_5$ appear in the positive and negative $y$ regions of the transverse plane, respectively.   However, the average change in the velocity of positively charged particles is the same as that of negatively charged particles with both given by $\Delta v=v\frac{e^{\mu_R/T}-e^{\mu_L/T}}{e^{\mu_R/T}+e^{\mu_L/T}}$, where $v$ is the magnitude of average modified velocity of particles of right or left helicity, and $\mu_R$ and $\mu_L$ are chemical potentials of right and left chiralities, respectively.  Negatively and positively charged particles thus have same spatial distributions if their initial spatial distributions are the same.  Therefore, no eccentricity difference between negatively and positively charged particle appears even though an additional charge quadrupole moment is generated in the transverse plane by the CVW.  The effect of CVW is thus different from that of CMW, which causes the average change in the velocities of negatively and positively charged particles to have opposite sign given by $\Delta v$ and $-\Delta v$, respectively, and can thus result in a larger eccentricity for negatively charged particles than for positively charged particles.

Our result obtained with only the vorticity field differs from that of Ref.~\cite{PhysRevD.92.071501} based on schematic considerations. Using the argument for the effect of the chiral magnetic wave introduced in Ref.~\cite{PhysRevLett.107.052303} based on consideration of the net charge distribution, it is argued in Ref.~\cite{PhysRevD.92.071501} that the additional charge quadrupole moment generated by the vorticity field would lead to an elliptic flow splitting between negatively and positively charged particles.  As discussed in the previous paragraph, since the spatial distributions of negatively and positively charged particles remains the same, the additional charge quadrupole moment does not lead to different eccentricities between negatively and positively charged particles and thus cannot generate a splitting between their elliptic flows. This is different from the case with only the magnetic field, where the additional quadrupole momentum generated by the chiral magnetic wave can indeed lead to different eccentricities and thus elliptic flows between negatively and positively charged particles as shown in our previous study~\cite{PhysRevC.94.045204}.

\begin{figure}[h]
\centering
\includegraphics[width=0.5\textwidth]{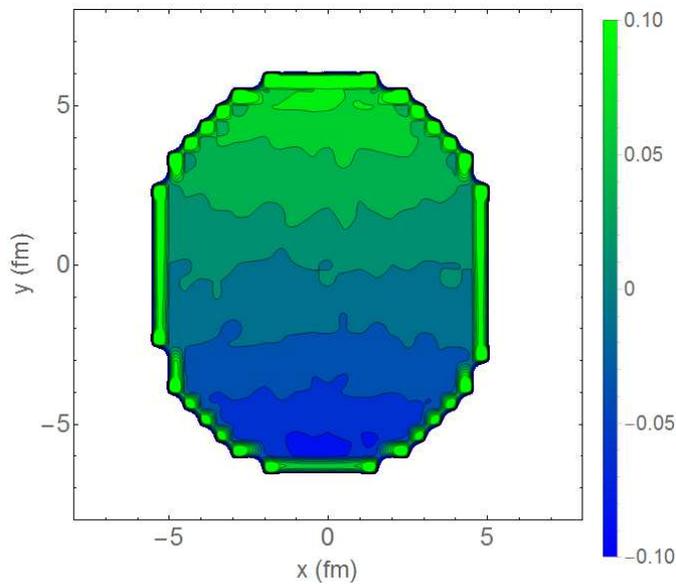}
\caption{(Color online) Axial charge chemical potential $\mu_5/T$ distribution in the transverse plane $z=0$ at time $t=5$ fm/$c$ for partonic matter of zero charge asymmetry.}
\label{dipole}
\end{figure}

Although the CVW does not lead to the eccentricity and elliptic flow splittings between negatively and positively charged particles, it does result in a large axial charge dipole moment in the transverse plane even for quark matter of vanishing charge asymmetry $A_{\pm}=0$.  As shown in Fig. \ref{dipole}, the value is even larger than that in
Ref.~\cite{PhysRevC.94.045204} due to the CMW based on a magnetic field of strength $eB=7m_{\pi}^2$.  To understand this, we note that the axial charge current induced by the magnetic and vorticity fields is~\cite{PhysRevLett.103.191601}
\begin{eqnarray}
\mathbf{j}_5\propto\left(\frac{1}{6}T^2+\frac{1}{2\pi^2}(\mu^2+\mu_5^2)\right)\boldsymbol{\omega}+\frac{Q}{2\pi^2}\mu\mathbf{B}.
\end{eqnarray}
With $T=0.3$ GeV, $\mu=\mu_5=0$, $\omega=0.02$ GeV in the present study, the axial charge current is 0.0375/fm$^3$, which is larger than the value 0.02/fm$^3$ obtained with $\mu/T=0.16$ and $QB=3.5m_{\pi}^2$ used in our previous study at same temperature~\cite{PhysRevC.94.045204}.

\subsection{\bf Effect of vorticity field plus magnetic field without the Lorentz force}

\begin{figure}[h]
\centering
\includegraphics[width=0.5\textwidth]{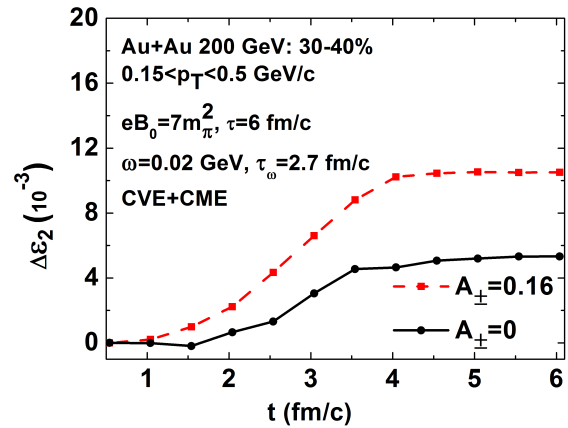}
\caption{(Color online) Eccentricity difference between negatively and positively charged particles as a function of time for partonic matter of charge asymmetries $A_{\pm}=0$ and $A_{\pm}=0.16$.}
\label{cmw1}
\end{figure}

Including also the magnetic field but still neglecting the Lorentz force in the chiral equations of motion, we have studied the time evolution of the difference between the eccentricities of negatively and positively charged particles for partonic matter of charge asymmetries $A_{\pm}=0$ and $A_{\pm}=0.16$.  Results for particles of momenta in the range $0.15\le p_T\le0.5$ GeV/$c$ are shown in Fig.~\ref{cmw1}. It is seen that the eccentricity difference increases with time in both cases and is nonzero even for zero charge asymmetry, indicating that effects of the vorticity field and the magnetic field are not additive because the eccentricity difference is zero in both cases for zero charge asymmetry. The latter is due to the finite axial charge current induced by the vorticity field, which leads to an axial charge dipole moment in the transverse plane, characterized by positive and negative axial charge chemical potentials in the positive and negative $y$ regions of the transverse plane, respectively.  Because of the chiral magnetic effect, there is a vector charge current along the magnetic field in the positive $y$ region and opposite to the magnetic field in the negative $y$ region. This then leads to a vector charge quadrupole moment in the transverse plane even when the charge asymmetry is zero.

\begin{figure}[h]
\centering
\includegraphics[width=0.5\textwidth]{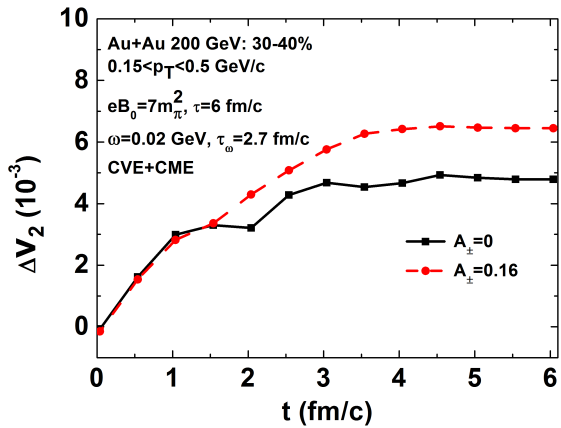}
\caption{(Color online) Same as Fig. \ref{cmw1} for the elliptic flow difference between positively and negatively charged particles as a function of time.}
\label{cmw2}
\end{figure}

We further show in Fig. \ref{cmw2} the time evolution of the elliptic flow difference between negatively and positively charged particles due to both the vorticity and the magnetic field but with the neglect of the Lorentz force for quark matter of charge asymmetries $A_{\pm}=0$ and $A_{\pm}=0.16$.  A finite positive elliptic flow difference is seen for zero charge asymmetry as a result of the finite eccentricity difference shown in Fig.~\ref{cmw1}.  Compared to the case of having only the magnetic field ~\cite{PhysRevC.94.045204}, the elliptic flow difference in the present case of having also the vorticity field appears earlier in time, and this is because the vorticity field helps to generate different average velocities for negatively and positively charged particles more quickly, which is in contrast to the case with only the magnetic field, where this difference is proportional to the axial charge chemical potential and thus takes time to develop.

\subsection{Effect of vorticity field plus magnetic field with the Lorentz force}

\begin{figure}[h]
\centering
\includegraphics[width=0.5\textwidth]{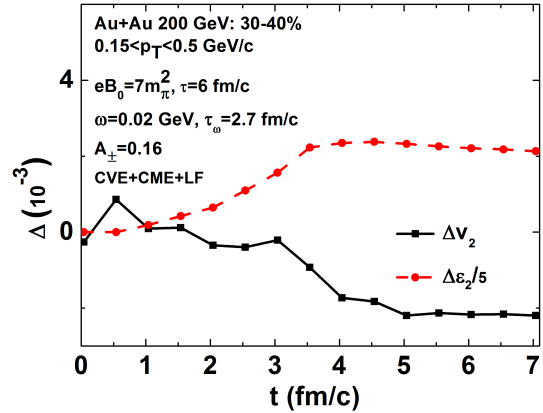}
\caption{(Color online) Eccentricity and elliptic flow differences between negatively and positively charged particles as a function of time for different scenarios of parton dynamics when the total charge asymmetry of the quark matter is $A_{\pm}=0.16$.}
\label{lorentz}
\end{figure}

Effect of the Lorentz force on the eccentricity and elliptic flow differences between negatively and positively charged particles is shown in Fig. \ref{lorentz} for partonic matter of charge asymmetry $A_{\pm}=0.16$. Compared to that shown in Fig. \ref{cmw1} without the Lorentz force, the eccentricity difference between negatively and positively charged particles with transverse momenta in the range $0.15\le p_T\le0.5$ GeV/$c$ is slightly larger after including the Lorentz force, and the difference decreases after reaching a maximum value.  For the time evolution of the elliptic flow difference between negatively and positively charged particles, Fig. \ref{lorentz} shows that the combined effects of the vorticity and magnetic fields with the inclusion of the Lorentz force lead to an initial increase of the elliptic flow difference, which then quickly decreases and becomes negative.  The Lorentz force thus cancels the chiral effects due to the vorticity and magnetic fields shown in Fig. \ref{cmw2} and even leads to an opposite effect on the elliptic flow difference between negatively and positively charged particles.

\subsection{Charge asymmetry dependence of the elliptic flow difference}

\begin{figure}[h]
\centering
\includegraphics[width=0.5\textwidth]{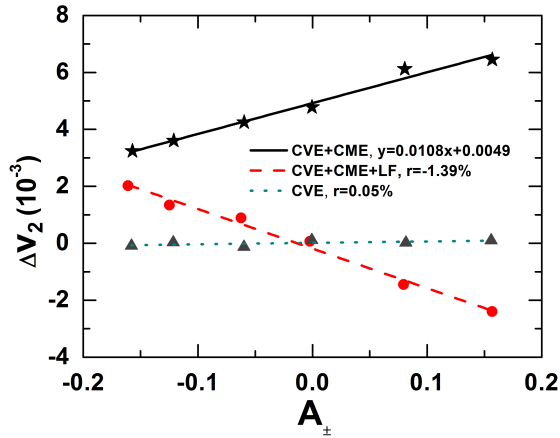}
\caption{(Color online) Elliptic flow difference as a function of charge asymmetry $A_{\pm}$ for different scenarios of parton dynamics.}
\label{elliptic}
\end{figure}

In Fig. \ref{elliptic}, we show the charge asymmetry dependence of the elliptic flow difference between negatively and positively charged particles for different scenarios of parton dynamics. The dotted line shows that the elliptic flow difference is consistent with zero for the case when only the vorticity field is present (CVE), and this is because the vorticity field has same effects on negatively and positively charged particles. When both the vorticity and the magnetic field are present but without including the Lorentz force (CVE+CME), the elliptic flow difference increases linearly with the charge asymmetry of the partonic matter with a slope parameter 0.0108, which is comparable to the results obtained in Refs.~\cite{PhysRevC.89.044909,PhysRevC.94.045204} based on the CMW. However, the inclusion of the vorticity field leads to a finite intercept of 4.9$\times 10^{-3}$ at zero charge asymmetry, and this is due to fact that the axial charge dipole moment in the transverse plane generated by the vorticity field can subsequently induce by the magnetic field a vector charge quadrupole moment of different eccentricities for negatively and positively charged particles. With the inclusion of the Lorentz force, which is shown by the dashed line, chiral effects due to the vorticity and magnetic fields disappear, and the slope parameter in the charge asymmetry dependence of the elliptic flow difference becomes negative with a magnitude of 0.0139.

\section{Summary}

Based on the anomalous transport model, which includes the propagation of massless quarks and antiquarks according to the chiral equations of motion and the chirality-changing scattering, we have studied the elliptic flows of charged particles in non-central heavy ion collisions at relativistic energies. Using initial conditions from a blast wave model and assuming the presence of only the vorticity field, which is modeled according to that from the AMPT model~\cite{Lin:2004en}, we find that the CVW does not lead to an elliptic flow splitting of negatively and positively charged particles when the charge asymmetry of the partonic matter is nonzero. On the other hand, including also a strong and long-lived magnetic field but neglecting the Lorentz force can lead to a splitting between the elliptic flows of negatively and positively charged particles.  However, the slope parameter in the charge asymmetry dependence of the elliptic flow splitting is smaller than the experimental data, while the positive intercept at zero charge asymmetry is larger than the experimental value~\cite{Wang2013248c}.   Unfortunately, as in our previous study including only the magnetic field~\cite{PhysRevC.94.045204}, the inclusion of the Lorentz force cancells the chiral effects due to the magnetic and vorticity fields and leads instead to a negative slope parameter in the charge symmetry dependence of the elliptic flow splitting of negatively and positively charged particles, contrary to that observed in experiments. Understanding this experimental results in terms of the chiral effects thus remains a challenge.  On the other hand, the large axial charge dipole moment generated in the transverse plane of a heavy ion collision by the CVW can lead to the polarization of particles with spins, which may be relevant to the observed finite polarization of lambda hyperons at RHIC~\cite{1742-6596-736-1-012016}. We plan to investigate this quantitatively in a future study.

\section*{Acknowledgements}

This work was supported in part by the US Department of Energy under Contract No. DE-SC0015266 and the Welch Foundation under Grant No. A-1358.

\bibliography{ref}

\begin{thebibliography}{32}
\expandafter\ifx\csname natexlab\endcsname\relax\def\natexlab#1{#1}\fi
\expandafter\ifx\csname bibnamefont\endcsname\relax
  \def\bibnamefont#1{#1}\fi
\expandafter\ifx\csname bibfnamefont\endcsname\relax
  \def\bibfnamefont#1{#1}\fi
\expandafter\ifx\csname citenamefont\endcsname\relax
  \def\citenamefont#1{#1}\fi
\expandafter\ifx\csname url\endcsname\relax
  \def\url#1{\texttt{#1}}\fi
\expandafter\ifx\csname urlprefix\endcsname\relax\def\urlprefix{URL }\fi
\providecommand{\bibinfo}[2]{#2}
\providecommand{\eprint}[2][]{\url{#2}}

\bibitem[{\citenamefont{Charbonneau and Zhitnitsky}(2010)}]{Charbonneau:2009ax}
\bibinfo{author}{\bibfnamefont{J.}~\bibnamefont{Charbonneau}} \bibnamefont{and}
  \bibinfo{author}{\bibfnamefont{A.}~\bibnamefont{Zhitnitsky}},
  \bibinfo{journal}{JCAP} \textbf{\bibinfo{volume}{1008}}, \bibinfo{pages}{010}
  (\bibinfo{year}{2010}).

\bibitem[{\citenamefont{Zyuzin and Burkov}(2012)}]{PhysRevB.86.115133}
\bibinfo{author}{\bibfnamefont{A.~A.} \bibnamefont{Zyuzin}} \bibnamefont{and}
  \bibinfo{author}{\bibfnamefont{A.~A.} \bibnamefont{Burkov}},
  \bibinfo{journal}{Phys. Rev. B} \textbf{\bibinfo{volume}{86}},
  \bibinfo{pages}{115133} (\bibinfo{year}{2012}).

\bibitem[{\citenamefont{Ba\ifmmode~\mbox{\c{s}}\else \c{s}\fi{}ar
  et~al.}(2014)\citenamefont{Ba\ifmmode~\mbox{\c{s}}\else \c{s}\fi{}ar,
  Kharzeev, and Yee}}]{PhysRevB.89.035142}
\bibinfo{author}{\bibfnamefont{G.~m.~c.}
  \bibnamefont{Ba\ifmmode~\mbox{\c{s}}\else \c{s}\fi{}ar}},
  \bibinfo{author}{\bibfnamefont{D.~E.} \bibnamefont{Kharzeev}},
  \bibnamefont{and} \bibinfo{author}{\bibfnamefont{H.-U.} \bibnamefont{Yee}},
  \bibinfo{journal}{Phys. Rev. B} \textbf{\bibinfo{volume}{89}},
  \bibinfo{pages}{035142} (\bibinfo{year}{2014}).

\bibitem[{\citenamefont{Son and Zhitnitsky}(2004)}]{PhysRevD.70.074018}
\bibinfo{author}{\bibfnamefont{D.~T.} \bibnamefont{Son}} \bibnamefont{and}
  \bibinfo{author}{\bibfnamefont{A.~R.} \bibnamefont{Zhitnitsky}},
  \bibinfo{journal}{Phys. Rev. D} \textbf{\bibinfo{volume}{70}},
  \bibinfo{pages}{074018} (\bibinfo{year}{2004}).

\bibitem[{\citenamefont{Metlitski and Zhitnitsky}(2005)}]{PhysRevD.72.045011}
\bibinfo{author}{\bibfnamefont{M.~A.} \bibnamefont{Metlitski}}
  \bibnamefont{and} \bibinfo{author}{\bibfnamefont{A.~R.}
  \bibnamefont{Zhitnitsky}}, \bibinfo{journal}{Phys. Rev. D}
  \textbf{\bibinfo{volume}{72}}, \bibinfo{pages}{045011}
  (\bibinfo{year}{2005}).

\bibitem[{\citenamefont{Kharzeev et~al.}(2008)\citenamefont{Kharzeev, McLerran,
  and Warringa}}]{Kharzeev2008227}
\bibinfo{author}{\bibfnamefont{D.~E.} \bibnamefont{Kharzeev}},
  \bibinfo{author}{\bibfnamefont{L.~D.} \bibnamefont{McLerran}},
  \bibnamefont{and} \bibinfo{author}{\bibfnamefont{H.~J.}
  \bibnamefont{Warringa}}, \bibinfo{journal}{Nuclear Physics A}
  \textbf{\bibinfo{volume}{803}}, \bibinfo{pages}{227 } (\bibinfo{year}{2008}).

\bibitem[{\citenamefont{Fukushima et~al.}(2008)\citenamefont{Fukushima,
  Kharzeev, and Warringa}}]{PhysRevD.78.074033}
\bibinfo{author}{\bibfnamefont{K.}~\bibnamefont{Fukushima}},
  \bibinfo{author}{\bibfnamefont{D.~E.} \bibnamefont{Kharzeev}},
  \bibnamefont{and} \bibinfo{author}{\bibfnamefont{H.~J.}
  \bibnamefont{Warringa}}, \bibinfo{journal}{Phys. Rev. D}
  \textbf{\bibinfo{volume}{78}}, \bibinfo{pages}{074033}
  (\bibinfo{year}{2008}).

\bibitem[{\citenamefont{Kharzeev}(2010)}]{Kharzeev2010205}
\bibinfo{author}{\bibfnamefont{D.~E.} \bibnamefont{Kharzeev}},
  \bibinfo{journal}{Annals of Physics} \textbf{\bibinfo{volume}{325}},
  \bibinfo{pages}{205 } (\bibinfo{year}{2010}).

\bibitem[{\citenamefont{Son and Sur\'owka}(2009)}]{PhysRevLett.103.191601}
\bibinfo{author}{\bibfnamefont{D.~T.} \bibnamefont{Son}} \bibnamefont{and}
  \bibinfo{author}{\bibfnamefont{P.}~\bibnamefont{Sur\'owka}},
  \bibinfo{journal}{Phys. Rev. Lett.} \textbf{\bibinfo{volume}{103}},
  \bibinfo{pages}{191601} (\bibinfo{year}{2009}).

\bibitem[{\citenamefont{Burnier et~al.}(2011)\citenamefont{Burnier, Kharzeev,
  Liao, and Yee}}]{PhysRevLett.107.052303}
\bibinfo{author}{\bibfnamefont{Y.}~\bibnamefont{Burnier}},
  \bibinfo{author}{\bibfnamefont{D.~E.} \bibnamefont{Kharzeev}},
  \bibinfo{author}{\bibfnamefont{J.}~\bibnamefont{Liao}}, \bibnamefont{and}
  \bibinfo{author}{\bibfnamefont{H.-U.} \bibnamefont{Yee}},
  \bibinfo{journal}{Phys. Rev. Lett.} \textbf{\bibinfo{volume}{107}},
  \bibinfo{pages}{052303} (\bibinfo{year}{2011}).

\bibitem[{\citenamefont{Jiang et~al.}(2015)\citenamefont{Jiang, Huang, and
  Liao}}]{PhysRevD.92.071501}
\bibinfo{author}{\bibfnamefont{Y.}~\bibnamefont{Jiang}},
  \bibinfo{author}{\bibfnamefont{X.-G.} \bibnamefont{Huang}}, \bibnamefont{and}
  \bibinfo{author}{\bibfnamefont{J.}~\bibnamefont{Liao}},
  \bibinfo{journal}{Phys. Rev. D} \textbf{\bibinfo{volume}{92}},
  \bibinfo{pages}{071501} (\bibinfo{year}{2015}).

\bibitem[{\citenamefont{Voronyuk et~al.}(2011)\citenamefont{Voronyuk, Toneev,
  Cassing, Bratkovskaya, Konchakovski, and Voloshin}}]{PhysRevC.83.054911}
\bibinfo{author}{\bibfnamefont{V.}~\bibnamefont{Voronyuk}},
  \bibinfo{author}{\bibfnamefont{V.~D.} \bibnamefont{Toneev}},
  \bibinfo{author}{\bibfnamefont{W.}~\bibnamefont{Cassing}},
  \bibinfo{author}{\bibfnamefont{E.~L.} \bibnamefont{Bratkovskaya}},
  \bibinfo{author}{\bibfnamefont{V.~P.} \bibnamefont{Konchakovski}},
  \bibnamefont{and} \bibinfo{author}{\bibfnamefont{S.~A.}
  \bibnamefont{Voloshin}}, \bibinfo{journal}{Phys. Rev. C}
  \textbf{\bibinfo{volume}{83}}, \bibinfo{pages}{054911}
  (\bibinfo{year}{2011}).

\bibitem[{\citenamefont{Deng and Huang}(2012)}]{PhysRevC.85.044907}
\bibinfo{author}{\bibfnamefont{W.-T.} \bibnamefont{Deng}} \bibnamefont{and}
  \bibinfo{author}{\bibfnamefont{X.-G.} \bibnamefont{Huang}},
  \bibinfo{journal}{Phys. Rev. C} \textbf{\bibinfo{volume}{85}},
  \bibinfo{pages}{044907} (\bibinfo{year}{2012}).

\bibitem[{\citenamefont{Stephanov and Yin}(2012)}]{PhysRevLett.109.162001}
\bibinfo{author}{\bibfnamefont{M.~A.} \bibnamefont{Stephanov}}
  \bibnamefont{and} \bibinfo{author}{\bibfnamefont{Y.}~\bibnamefont{Yin}},
  \bibinfo{journal}{Phys. Rev. Lett.} \textbf{\bibinfo{volume}{109}},
  \bibinfo{pages}{162001} (\bibinfo{year}{2012}).

\bibitem[{\citenamefont{Son and Yamamoto}(2012)}]{Son:2012wh}
\bibinfo{author}{\bibfnamefont{D.~T.} \bibnamefont{Son}} \bibnamefont{and}
  \bibinfo{author}{\bibfnamefont{N.}~\bibnamefont{Yamamoto}},
  \bibinfo{journal}{Phys. Rev. Lett.} \textbf{\bibinfo{volume}{109}},
  \bibinfo{pages}{181602} (\bibinfo{year}{2012}).

\bibitem[{\citenamefont{Son and Yamamoto}(2013)}]{Son:2012zy}
\bibinfo{author}{\bibfnamefont{D.~T.} \bibnamefont{Son}} \bibnamefont{and}
  \bibinfo{author}{\bibfnamefont{N.}~\bibnamefont{Yamamoto}},
  \bibinfo{journal}{Phys. Rev.} \textbf{\bibinfo{volume}{D87}},
  \bibinfo{pages}{085016} (\bibinfo{year}{2013}).

\bibitem[{\citenamefont{Gao et~al.}(2012)\citenamefont{Gao, Liang, Pu, Wang,
  and Wang}}]{PhysRevLett.109.232301}
\bibinfo{author}{\bibfnamefont{J.-H.} \bibnamefont{Gao}},
  \bibinfo{author}{\bibfnamefont{Z.-T.} \bibnamefont{Liang}},
  \bibinfo{author}{\bibfnamefont{S.}~\bibnamefont{Pu}},
  \bibinfo{author}{\bibfnamefont{Q.}~\bibnamefont{Wang}}, \bibnamefont{and}
  \bibinfo{author}{\bibfnamefont{X.-N.} \bibnamefont{Wang}},
  \bibinfo{journal}{Phys. Rev. Lett.} \textbf{\bibinfo{volume}{109}},
  \bibinfo{pages}{232301} (\bibinfo{year}{2012}).

\bibitem[{\citenamefont{Chen et~al.}(2013)\citenamefont{Chen, Pu, Wang, and
  Wang}}]{PhysRevLett.110.262301}
\bibinfo{author}{\bibfnamefont{J.-W.} \bibnamefont{Chen}},
  \bibinfo{author}{\bibfnamefont{S.}~\bibnamefont{Pu}},
  \bibinfo{author}{\bibfnamefont{Q.}~\bibnamefont{Wang}}, \bibnamefont{and}
  \bibinfo{author}{\bibfnamefont{X.-N.} \bibnamefont{Wang}},
  \bibinfo{journal}{Phys. Rev. Lett.} \textbf{\bibinfo{volume}{110}},
  \bibinfo{pages}{262301} (\bibinfo{year}{2013}).

\bibitem[{\citenamefont{Manuel and Torres-Rincon}(2014)}]{Manuel:2014dza}
\bibinfo{author}{\bibfnamefont{C.}~\bibnamefont{Manuel}} \bibnamefont{and}
  \bibinfo{author}{\bibfnamefont{J.~M.} \bibnamefont{Torres-Rincon}},
  \bibinfo{journal}{Phys. Rev.} \textbf{\bibinfo{volume}{D90}},
  \bibinfo{pages}{076007} (\bibinfo{year}{2014}).

\bibitem[{\citenamefont{Chen et~al.}(2014)\citenamefont{Chen, Son, Stephanov,
  Yee, and Yin}}]{PhysRevLett.113.182302}
\bibinfo{author}{\bibfnamefont{J.-Y.} \bibnamefont{Chen}},
  \bibinfo{author}{\bibfnamefont{D.~T.} \bibnamefont{Son}},
  \bibinfo{author}{\bibfnamefont{M.~A.} \bibnamefont{Stephanov}},
  \bibinfo{author}{\bibfnamefont{H.-U.} \bibnamefont{Yee}}, \bibnamefont{and}
  \bibinfo{author}{\bibfnamefont{Y.}~\bibnamefont{Yin}},
  \bibinfo{journal}{Phys. Rev. Lett.} \textbf{\bibinfo{volume}{113}},
  \bibinfo{pages}{182302} (\bibinfo{year}{2014}).

\bibitem[{\citenamefont{Chen et~al.}(2015)\citenamefont{Chen, Son, and
  Stephanov}}]{PhysRevLett.115.021601}
\bibinfo{author}{\bibfnamefont{J.-Y.} \bibnamefont{Chen}},
  \bibinfo{author}{\bibfnamefont{D.~T.} \bibnamefont{Son}}, \bibnamefont{and}
  \bibinfo{author}{\bibfnamefont{M.~A.} \bibnamefont{Stephanov}},
  \bibinfo{journal}{Phys. Rev. Lett.} \textbf{\bibinfo{volume}{115}},
  \bibinfo{pages}{021601} (\bibinfo{year}{2015}).

\bibitem[{\citenamefont{Sun et~al.}(2016)\citenamefont{Sun, Ko, and
  Li}}]{PhysRevC.94.045204}
\bibinfo{author}{\bibfnamefont{Y.}~\bibnamefont{Sun}},
  \bibinfo{author}{\bibfnamefont{C.~M.} \bibnamefont{Ko}}, \bibnamefont{and}
  \bibinfo{author}{\bibfnamefont{F.}~\bibnamefont{Li}}, \bibinfo{journal}{Phys.
  Rev. C} \textbf{\bibinfo{volume}{94}}, \bibinfo{pages}{045204}
  (\bibinfo{year}{2016}).

\bibitem[{\citenamefont{Ba\ifmmode~\mbox{\c{s}}\else \c{s}\fi{}ar
  et~al.}(2012)\citenamefont{Ba\ifmmode~\mbox{\c{s}}\else \c{s}\fi{}ar,
  Kharzeev, and Skokov}}]{PhysRevLett.109.202303}
\bibinfo{author}{\bibfnamefont{G.}~\bibnamefont{Ba\ifmmode~\mbox{\c{s}}\else
  \c{s}\fi{}ar}}, \bibinfo{author}{\bibfnamefont{D.~E.}
  \bibnamefont{Kharzeev}}, \bibnamefont{and}
  \bibinfo{author}{\bibfnamefont{V.}~\bibnamefont{Skokov}},
  \bibinfo{journal}{Phys. Rev. Lett.} \textbf{\bibinfo{volume}{109}},
  \bibinfo{pages}{202303} (\bibinfo{year}{2012}).

\bibitem[{\citenamefont{Jiang et~al.}(2016)\citenamefont{Jiang, Lin, and
  Liao}}]{PhysRevC.94.044910}
\bibinfo{author}{\bibfnamefont{Y.}~\bibnamefont{Jiang}},
  \bibinfo{author}{\bibfnamefont{Z.-W.} \bibnamefont{Lin}}, \bibnamefont{and}
  \bibinfo{author}{\bibfnamefont{J.}~\bibnamefont{Liao}},
  \bibinfo{journal}{Phys. Rev. C} \textbf{\bibinfo{volume}{94}},
  \bibinfo{pages}{044910} (\bibinfo{year}{2016}).

\bibitem[{\citenamefont{Lin et~al.}(2005)\citenamefont{Lin, Ko, Li, Zhang, and
  Pal}}]{Lin:2004en}
\bibinfo{author}{\bibfnamefont{Z.-W.} \bibnamefont{Lin}},
  \bibinfo{author}{\bibfnamefont{C.~M.} \bibnamefont{Ko}},
  \bibinfo{author}{\bibfnamefont{B.-A.} \bibnamefont{Li}},
  \bibinfo{author}{\bibfnamefont{B.}~\bibnamefont{Zhang}}, \bibnamefont{and}
  \bibinfo{author}{\bibfnamefont{S.}~\bibnamefont{Pal}},
  \bibinfo{journal}{Phys. Rev.} \textbf{\bibinfo{volume}{C72}},
  \bibinfo{pages}{064901} (\bibinfo{year}{2005}).

\bibitem[{\citenamefont{Ghosh et~al.}(2016)\citenamefont{Ghosh, Peixoto, Roy,
  Serna, and Krein}}]{Ghosh:2015mda}
\bibinfo{author}{\bibfnamefont{S.}~\bibnamefont{Ghosh}},
  \bibinfo{author}{\bibfnamefont{T.~C.} \bibnamefont{Peixoto}},
  \bibinfo{author}{\bibfnamefont{V.}~\bibnamefont{Roy}},
  \bibinfo{author}{\bibfnamefont{F.~E.} \bibnamefont{Serna}}, \bibnamefont{and}
  \bibinfo{author}{\bibfnamefont{G.}~\bibnamefont{Krein}},
  \bibinfo{journal}{Phys. Rev.} \textbf{\bibinfo{volume}{C93}},
  \bibinfo{pages}{045205} (\bibinfo{year}{2016}).

\bibitem[{\citenamefont{Heinz and Snellings}(2013)}]{Heinz:2013th}
\bibinfo{author}{\bibfnamefont{U.}~\bibnamefont{Heinz}} \bibnamefont{and}
  \bibinfo{author}{\bibfnamefont{R.}~\bibnamefont{Snellings}},
  \bibinfo{journal}{Ann. Rev. Nucl. Part. Sci.} \textbf{\bibinfo{volume}{63}},
  \bibinfo{pages}{123} (\bibinfo{year}{2013}).

\bibitem[{\citenamefont{Li and Ko}(1995)}]{Li:1995pra}
\bibinfo{author}{\bibfnamefont{B.-A.} \bibnamefont{Li}} \bibnamefont{and}
  \bibinfo{author}{\bibfnamefont{C.~M.} \bibnamefont{Ko}},
  \bibinfo{journal}{Phys. Rev.} \textbf{\bibinfo{volume}{C52}},
  \bibinfo{pages}{2037} (\bibinfo{year}{1995}).

\bibitem[{\citenamefont{Adams et~al.}(2005)\citenamefont{Adams, Aggarwal,
  Ahammed, Amonett, Anderson, Arkhipkin, Averichev, Badyal, Bai, Balewski
  et~al.}}]{PhysRevC.72.014904}
\bibinfo{author}{\bibfnamefont{J.}~\bibnamefont{Adams}},
  \bibinfo{author}{\bibfnamefont{M.~M.} \bibnamefont{Aggarwal}},
  \bibinfo{author}{\bibfnamefont{Z.}~\bibnamefont{Ahammed}},
  \bibinfo{author}{\bibfnamefont{J.}~\bibnamefont{Amonett}},
  \bibinfo{author}{\bibfnamefont{B.~D.} \bibnamefont{Anderson}},
  \bibinfo{author}{\bibfnamefont{D.}~\bibnamefont{Arkhipkin}},
  \bibinfo{author}{\bibfnamefont{G.~S.} \bibnamefont{Averichev}},
  \bibinfo{author}{\bibfnamefont{S.~K.} \bibnamefont{Badyal}},
  \bibinfo{author}{\bibfnamefont{Y.}~\bibnamefont{Bai}},
  \bibinfo{author}{\bibfnamefont{J.}~\bibnamefont{Balewski}},
  \bibnamefont{et~al.} (\bibinfo{collaboration}{STAR and STAR-RICH
  Collaborations}), \bibinfo{journal}{Phys. Rev. C}
  \textbf{\bibinfo{volume}{72}}, \bibinfo{pages}{014904}
  (\bibinfo{year}{2005}).

\bibitem[{\citenamefont{Yee and Yin}(2014)}]{PhysRevC.89.044909}
\bibinfo{author}{\bibfnamefont{H.-U.} \bibnamefont{Yee}} \bibnamefont{and}
  \bibinfo{author}{\bibfnamefont{Y.}~\bibnamefont{Yin}},
  \bibinfo{journal}{Phys. Rev. C} \textbf{\bibinfo{volume}{89}},
  \bibinfo{pages}{044909} (\bibinfo{year}{2014}).

\bibitem[{\citenamefont{Wang}(2013)}]{Wang2013248c}
\bibinfo{author}{\bibfnamefont{G.}~\bibnamefont{Wang}}, \bibinfo{journal}{Nuc.
  Phys. A} \textbf{\bibinfo{volume}{904--905}}, \bibinfo{pages}{248c }
  (\bibinfo{year}{2013}).

\bibitem[{\citenamefont{Upsal}(2016)}]{1742-6596-736-1-012016}
\bibinfo{author}{\bibfnamefont{I.}~\bibnamefont{Upsal}},
  \bibinfo{journal}{Journal of Physics: Conference Series}
  \textbf{\bibinfo{volume}{736}}, \bibinfo{pages}{012016}
  (\bibinfo{year}{2016}).

\end{thebibliography}
\end{document}